\documentclass[12pt]{article}
\usepackage[utf8]{inputenc}
\usepackage{setspace}
\usepackage{amsmath}
\usepackage{graphics}
\usepackage{multicol}
\usepackage{subcaption}
\usepackage{graphicx}
\usepackage{placeins}
\usepackage{amsmath}
\usepackage{pdfpages}
\usepackage{graphicx}
\usepackage{xcolor}
\usepackage[labelfont=bf,tableposition=top]{caption}
\usepackage{blindtext}
\usepackage{tcolorbox}
\usepackage{amsthm}
\usepackage{lipsum}
\usepackage{multirow}
\usepackage{tabularray}
\usepackage{floatrow}

\newfloatcommand{capbtabbox}{table}[][\FBwidth]

\title{A flexible framework for synthesizing human activity patterns with application to sequential categorical data}
\author{Zuofu Huang, Julian Wolfson, Jayne A. Fulkerson, Ryan Demmer, Helen N. Chen}
\date{}

\begin{document}

\maketitle

\vspace{-5mm}

\begin{abstract}
    The ability to synthesize realistic data in a parametrizable way is valuable for a number of reasons, including privacy, missing data imputation, and evaluating the performance of statistical and computational methods. When the underlying data generating process is complex, data synthesis requires approaches that balance realism and simplicity. In this paper, we address the problem of synthesizing sequential categorical data of the type that is increasingly available from mobile applications and sensors that record participant status continuously over the course of multiple days and weeks. We propose the paired Markov Chain (paired-MC) method, a flexible framework that produces sequences that closely mimic real data while providing a straightforward mechanism for modifying characteristics of the synthesized sequences. We demonstrate the paired-MC method on two datasets, one reflecting daily human activity patterns collected via a smartphone application, and one encoding the intensities of physical activity measured by wearable accelerometers. In both settings, sequences synthesized by paired-MC better capture key characteristics of the real data than alternative approaches.
\end{abstract}


\section{Introduction}

Over the last few decades, an increasing amount of data are being collected using mobile and wearable devices which enable the recording of physical and situational measures on a near continuous basis with relatively low participant burden. For example, mobile apps can use contextual data on location, speed, etc. to infer human activity episodes (home, work, car, bus, etc.); the resulting data can be applied in areas from urban planning to public health. The motivating dataset comes from a study of healthcare workers' behaviors and risk factors associated with COVID-19. Data were collected using Daynamica, a mobile application that captures daily activities using smartphone sensors and statistical learning methods. \cite{fan2015smartrac} Our source dataset includes 1929 days from 68 individuals. On each day, the start and end time of that day's activities were automatically recorded by the Daynamica app, yielding a sequence of daily episodes labeled with one of 12 activity types, 13 if accounting for missing as a state (see Figure 1). 

\begin{figure}[h!]
    \centering
    \includegraphics[width = 140mm]{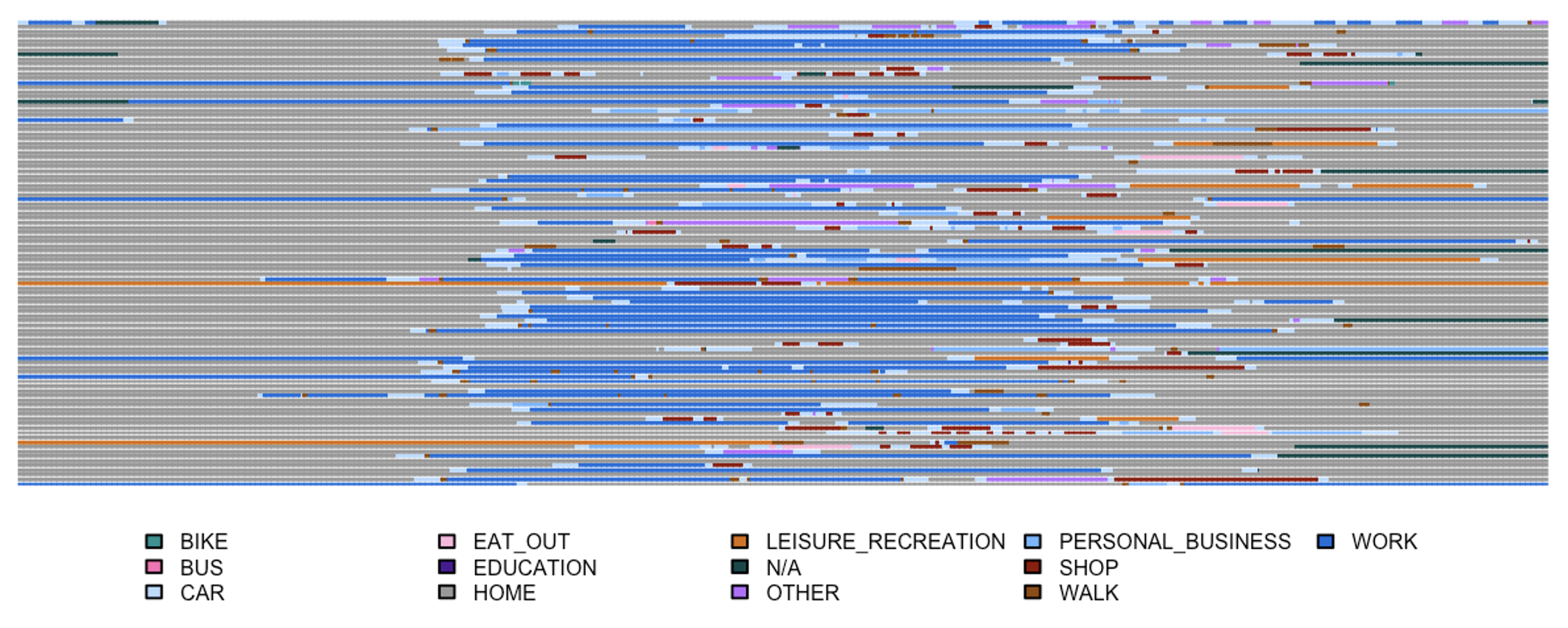} (a)

    \includegraphics[width = 140mm]{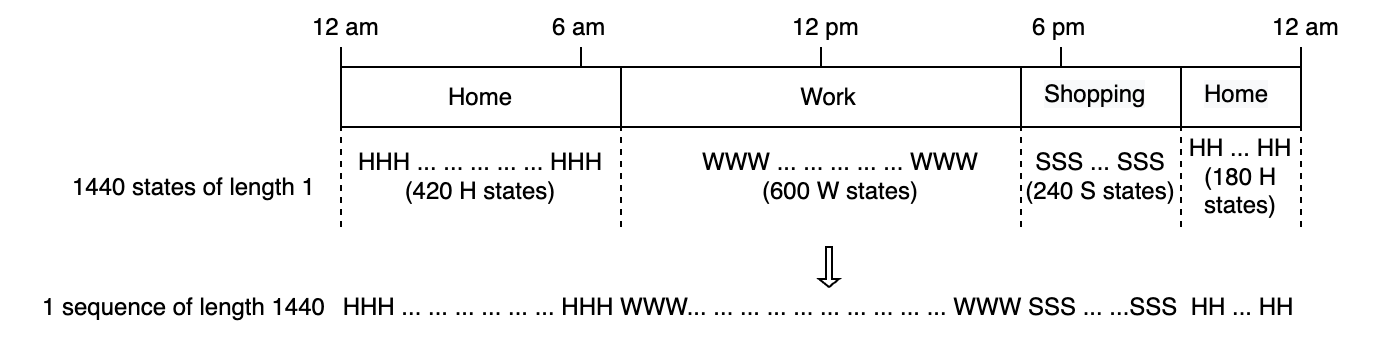} (b) \\

    \caption{(a) A sample of 100 daily sequences collected from healthcare workers (b) Characterization of an example daily sequence}
\end{figure}

While many representations of these episodic data are possible, we choose to encode them as a sequence of equal-duration intervals representing the current status during that interval. Under this representation, each day in our motivating data can be viewed as as a discrete-valued vector of 1440 1-minute intervals. \cite{song2021visualizing} Depending on the context, the duration of the intervals may be changed to reflect the frequency of sampling from our data collection process. Intensively sampled continuous-valued observations can  also be transformed into this representation by discretization; indeed, our second data application involves physical activity intensity data that are originally recorded as a continuous measure (steps per minute).

\subsection{Value of synthetic data}

The goal of this paper is to introduce a flexible, data-driven framework for synthesizing de novo sequential categorical data. There are a range of scenarios where synthesizing sequential categorical data would be beneficial. First, we may need to simulate data to better understand the performance of statistical approaches across a range of hypothetical underlying data generating processes. \cite{dahmen2019synsys} Simulated data should be realistic, in our case obeying the complex constraints of human activity sequences. Realism is not the only requirement for a synthesis framework, however; it should also be straightforward to manipulate the distribution of the synthesized data so that methods can be evaluated across multiple well-defined scenarios.  Second, synthetic data generation is an essential element of missing data imputation methods, a particularly acute need when dealing with automated data capture systems which can fail to record periods of time for a variety of reasons. Last, synthetic data may be preferred over real data due to concern of confidentiality or reidentification in many situations where the data provide detailed records about participants such as their locations and activities. \cite{benarous2022synthesis, bindschaedler2016synthesizing}  Our focus in this manuscript is mainly on describing our framework, showing that it generates realistic sequences, and demonstrating how the synthesis process can be tuned to produce sets of synthetic sequences with different characteristics. In the Discussion, we address how our framework might be applied to imputation and data privacy problems in the future.

\subsection{Approaches to synthesizing sequential categorical data}

Existing literature for synthesizing sequential categorical data has focused on synthesizing life trajectories in social sciences and human location sequence data. \cite{benarous2022synthesis, gabadinho2016analyzing} Generally, these sequences contain a relatively small number of sequential states, and lack the cyclic structure present in our data. We briefly review previous work on sequence synthesis, and discuss why they are not optimal for our data. 

\subsubsection{Observation-level synthesis methods}

In our representation, each sequence is encoded as a vector of observations indexed by their position in the sequence. Generally, synthesis approaches which operate directly on these position-specific observations will fail to capture the complex structure of the sequence. For example, the observation-level bootstrap and other resampling methods are unlikely to generate realistic sequences. The same limitation applies to methods such as genetic algorithms; small "local" manipulations (e.g., inserting an isolated "home" segment in the middle of a work day) are much more likely to result in implausible than plausible sequences.



\subsubsection{First- and higher-order Markov models}

Markov chains are canonical tools for modeling and synthesizing sequential categorical data. \cite{benarous2022synthesis, gabadinho2016analyzing} Given the nature of human activity sequences, the transition matrix shifts substantially by time of day, so the simplest Markov models using one universal transition matrix are inappropriate. For instance, the probability of having a work episode following a car episode is more likely in the morning than in the evening. Instead, we focus our attention on time-varying Markov Chains (TVMC), which provide reasonable and non-trivial comparators for our methods. Next, we discuss the respective reasons that make both continuous and discrete time-varying Markov chains unsuitable for modeling our data. 

\begin{figure}[h!]
    \centering
    \begin{multicols}{3}[]
        \includegraphics[width=42mm]{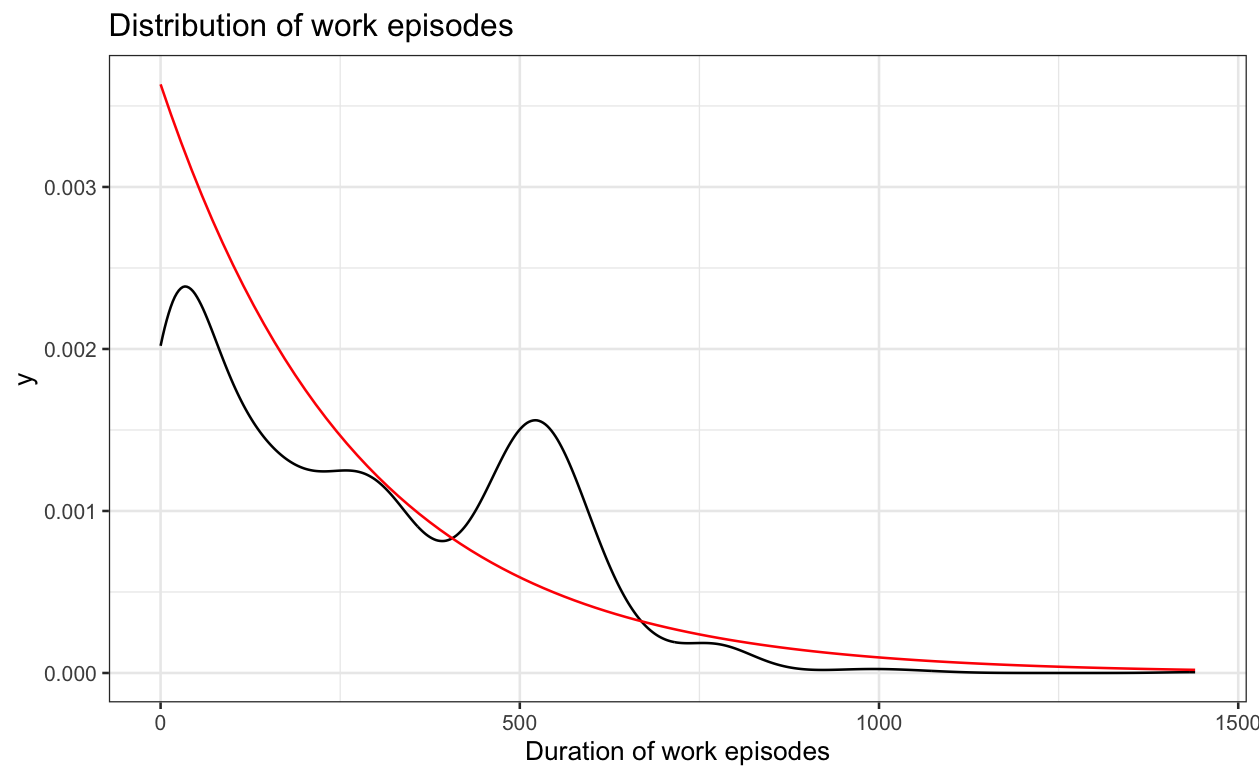}
        \includegraphics[width=42mm]{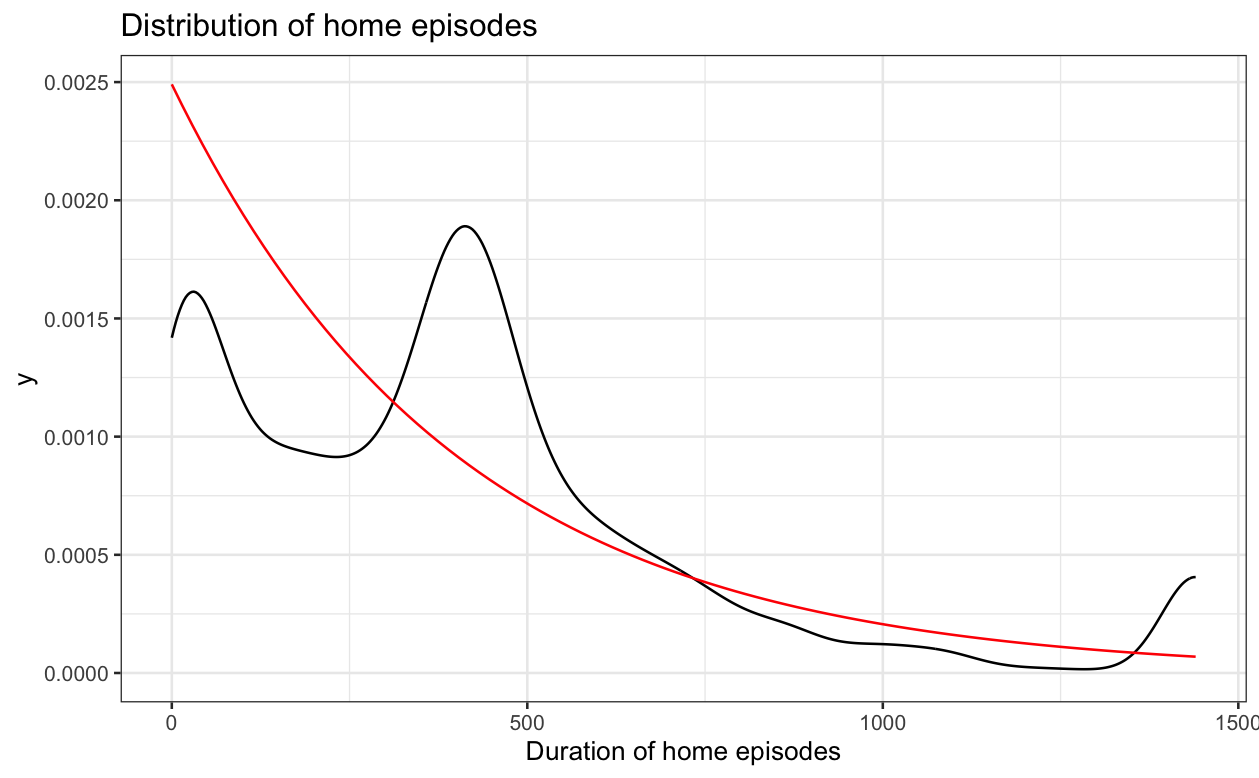}
        \includegraphics[width=42mm]{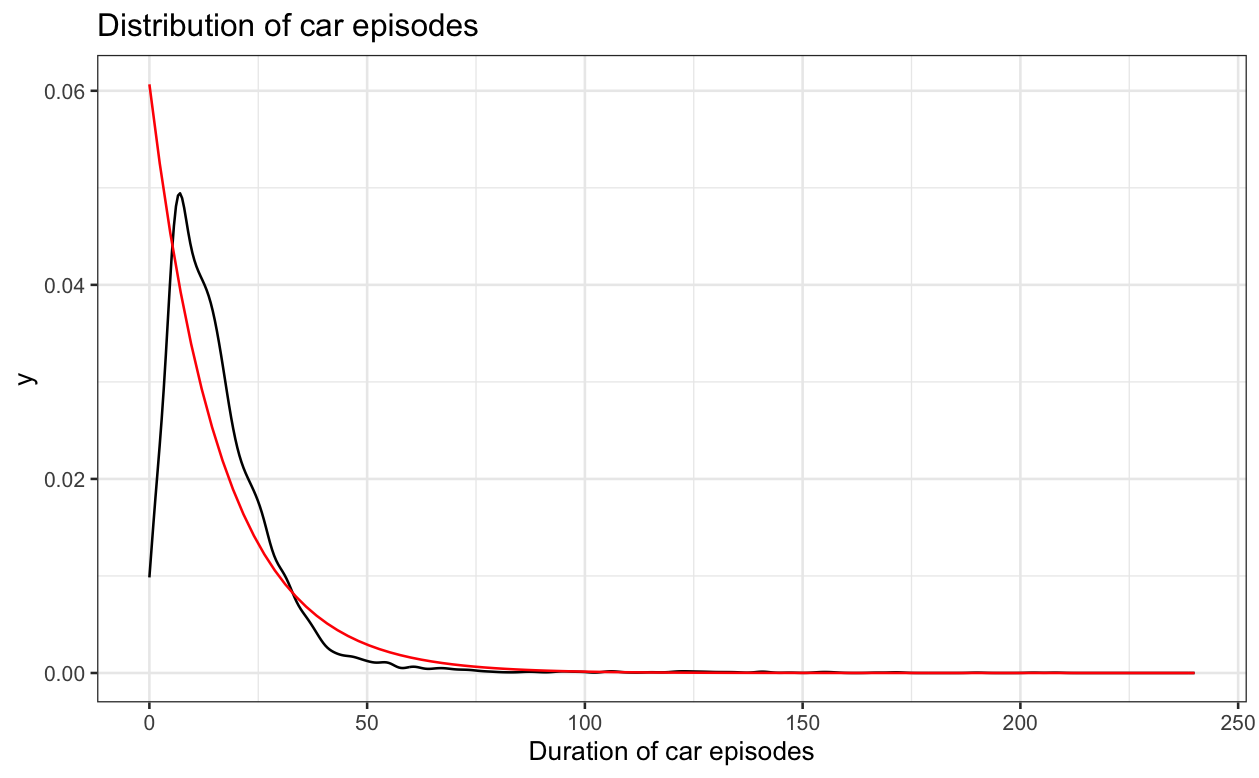}
    \end{multicols}
    \caption{Empirical distributions and best-fit exponential curves of work, home and car (0-4 hours) episode durations (from left to right)}
    \label{not_exponential}
\end{figure}

For continuous-time Markov Chains, the length of each state (i.e. holding time before transitioning into another state) needs to be exponentially distributed for the Markovian property to hold, since the exponential distribution is the only memoryless continuous distribution. However, state durations in our data are clearly not exponentially distributed (see Figure \ref{not_exponential}), so human activity sequences cannot be viewed as a non-homogeneous continuous-time Markov Chain.

For discrete-time Markov Chains, the Markovian property is clearly violated since the past state of the random process provides non-trivial information about the future. For example, a car episode to work at 8 am may indicate that the person is likely to drive home from work hours later (and thus several hundred states) during the day. Further problems will arise depending on whether the transition probability of the Markov Chain is estimated parametrically or non-parametrically. For parametric implementations of TVMC, prior knowledge of the relationship between transition probabilities and the time variable is required. This relationship is often unknown and difficult to estimate, where we need to know such relationships for each activity state while accounting for the correlation between them. For non-parametric implementations of TVMC, transitions can only happen if transitions are observed at the exact time, which does not offer adequate flexibility.

Markov Chains of higher orders (including variable-order Markov Chains) are often implemented to account for higher-order correlations. Probabilistic suffix trees (PST) have been implemented to store variable-length Markov chains and synthesize sequences based in the context of social sciences  \cite{gabadinho2016analyzing}. However, the order needs to be very high in longer sequences because the probability of transition between each step is often very low. First, this brings computational challenges as the run time is on the order of $O(n^k)$, where $n$ is the sequence length and $k$ is the order of the Markov Chain. It is difficult to choose the maximum depth of the PST ($L$) and the required minimal frequency of a subsequence to be added to the tree ($nmin$) that builds a computationally feasible PST without losing minute-level details of the sequence. Second, Markov Chains of higher orders are prone to overfitting since higher orders indicate increasingly specific patterns. An unrealistic amount of data are needed to capture higher-order patterns and the associated variability.

\subsubsection{Neural networks}

Deep learning techniques are increasingly popular in predicting time series and trajectories. While generative adversarial networks are commonly implemented in synthetic data generation, long short-term memory (LSTM) is more desirable for sequential data and has been implemented in scenarios both with and without spatial-temporal considerations \cite{benarous2022synthesis, alahi2016social, altche2017lstm}. Benarous et. al implemented LSTM to synthesize daily human location sequences with 18 states in each sequence. \cite{benarous2022synthesis} However, results showed that LSTM-based approaches perform comparatively worse than Markov-based methods in capturing the distribution of specific states. This result informs our hesitancy of using LSTM in synthesizing much longer sequences (such as human activity sequences with 1440 states given 1-minute sampling intervals). In addition, LSTM typically requires very large amounts of data to learn the complex structure of long sequences, and it may be computationally infeasible to synthesize a desired number of sequences. Finally, with LSTM, the user of the synthesis method has little control over the characteristics of the outputs that are produced; for example, shifting the distribution of generated sequences might require training an entirely new network.

\subsection{The paired-MC approach}

Our paired-MC method to sequence synthesis combines the characteristics of discrete-time and continuous-time Markov Chains and relaxes the Markovian property assumption. We further propose pre-clustering before synthesis that uncovers the subgroup structure within data, improves the accuracy of transition probability estimations and offers more flexible control of the output by reweighting each sequence. Synthesis results show that paired-MC is able to capture the true distribution of sequence states more accurately than Markov-based approaches, while maintaining other desirable statistical properties. While we motivate and illustrate the paired-MC method with sequential human activity data, we also demonstrate its application to NHANES accelerometer data.

The remainder of the article proceeds as follows. Section 2 describes our proposed synthesis approaches. Section 3 offers examples of the implementation and performance of our method. While we motivate and illustrate the paired-MC method with sequential human activity data, we also demonstrate its application to NHANES accelerometer data. We conclude with a discussion in Section 4.

\section{Methods}

\subsection{Proposed characterization of sequences: Paired-MC}

Consider a sequence of categorical states as state-duration pairs $\{(X_1, t_1), $ $(X_2, t_2), \dots, (X_n, t_n) \}$, where $n$ denotes the number of distinct episodes, $X_i$ the state at the $i$th distinct episode and $t_i$ the length of the corresponding state. We propose characterizing the sequence as a stochastic process $\mathcal{X} = \{(X_1 \rightarrow t_1) \rightarrow \dots \rightarrow (X_n \rightarrow t_n) \}$ where each state $X_i$ is paired with the time it spends in the state $t_i$ at each $(X_i \rightarrow t_i)$ step. (See Figure 3)

\begin{figure}[h!]
    \centering
    \includegraphics[width=150mm]{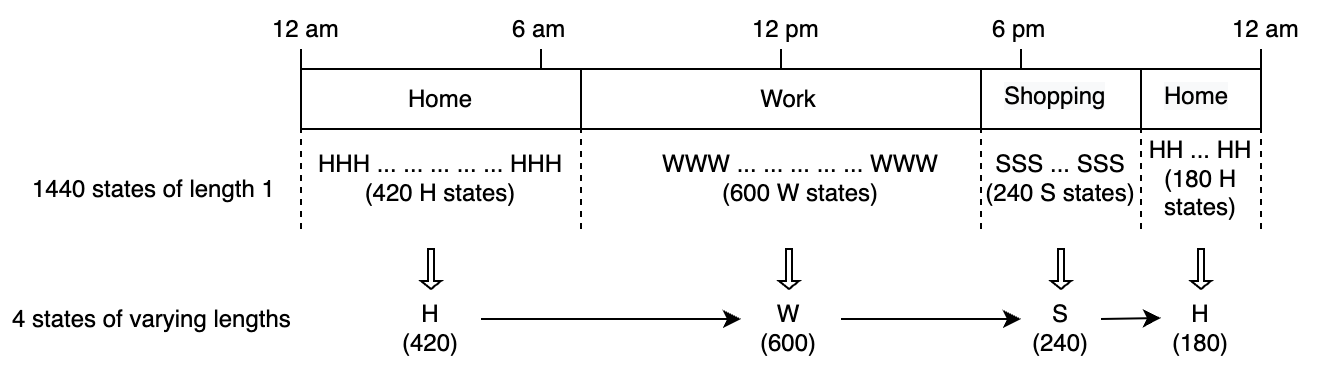}
    \caption{Paired-MC characterization of an example daily sequence}
    \label{pairedMC_example}
\end{figure}

Assuming that some level of dependence exists between adjacent states regardless of state lengths, $\{X_1 \rightarrow \dots \rightarrow X_n \}$ can be viewed as a discrete Markov Chain, where each "step" has varying lengths $t_i$ that does not have to be exponentially distributed. Such re-characterization bypasses the exponential constraint of continuous-time Markov Chains described in Section 1.2.2,
and effectively reduces a long Markov Chain to a stochastic process of the length equal to the number of distinct states in the sequence. This offers the chance to use tools such as Markov chains with order $k > 1$ (i.e. $p(X_i|X_{i-1}, \dots, X_{1}) = p(X_i |X_{i-1}, \dots, X_{i-k})$) to account for dependencies more accurately. 

\begin{figure}[h!]
    \centering
    \includegraphics[width=115mm]{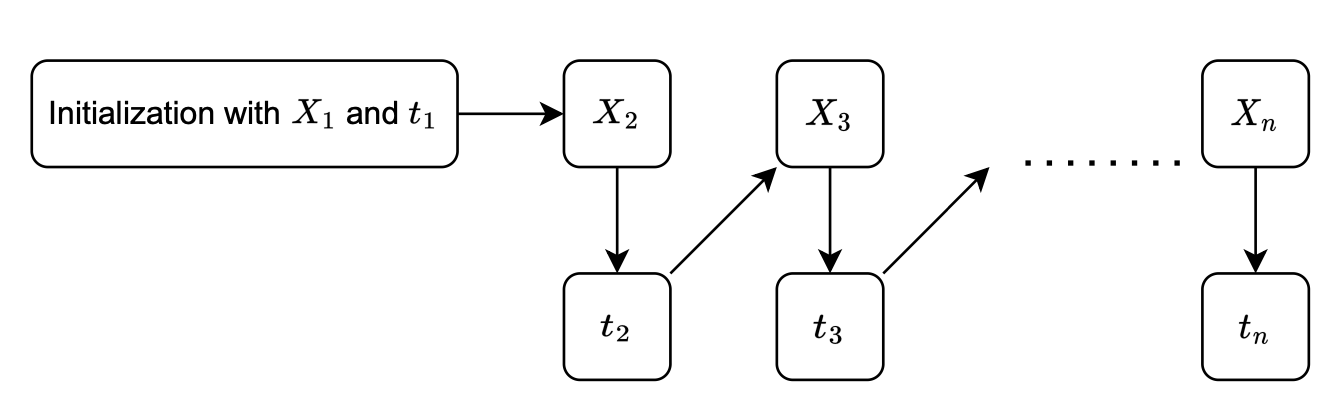}
    \caption{Visual representation of the paired-MC synthesis process}
    \label{process}
\end{figure}

\subsection{Implementation of Paired-MC}

Below we propose two synthesis methods that build on the characterization above: paired-MC of order 1 and paired-MC of higher orders.

\subsubsection*{Paired-MC of order 1}

To start, a random state is sampled according to the observed probabilities of occurrence from all observed states at time 1. The length of the first state is sampled from the distribution of all observed lengths of that state type that initiates at time 1. After sequence initialization, we generate the next state followed by its length, before we generate the next state. Lengths are determined in a separate process from the state. See Figure \ref{process}. Transition probability into the next state at each time $t$ is estimated by the observed transitions within a time window $\delta$ of $t$, say 1 hour for human activity sequences.  

There are many ways to obtain a possible length from the empirical distribution of observed lengths. Some strategies include sampling directly from the observations, from kernel density estimation (KDE) and from logspline density estimation. \cite{parzen1962estimation, kooperberg1992logspline, stone1997polynomial, silverman1986density} KDE and logspline estimation are initially considered due to privacy concerns over using observed state lengths. However, our simulation shows minimal concern for direct sampling due to different starting times and the abundance of potential lengths to choose from. Therefore, our discussion will focus on direct sampling, with KDE and logspline implemented as extensions.

Human activity sequence data are naturally organized into 24-hour periods corresponding to each day. Because the transition probability takes into account all transitions in a given time window $\delta$, the distribution of activities toward the end of sequence $(n-\delta, n)$ is biased. Therefore, we propose a small modification to the process, where we impute an additional length of $\delta$ as a buffer with easily implemented strategies such as TVMC, synthesize sequences of length $n+\delta$, and only take the first $n$ entries as the final result.

\vspace{7mm}

\framebox{\parbox{\dimexpr\linewidth-2\fboxsep-2\fboxrule}{\itshape
{\bf{Algorithm for paired-MC of order 1:}}
\begin{enumerate}
    \item At $t = 1$, initialize with a state and its length.
    \item Denote the current state as $A_c$ and the end time of $A_c$ as $t_c$. While $t_c < 1440$, repeat:
    \begin{enumerate}
        \item In a certain time interval $\delta$ of $t_c$ (e.g. $\delta = \pm$ 1 hour), find all states with $A_c$ as the preceding activity and their lengths. Each sequence may have more than 1 candidate that fulfills the requirement. 
        \item Randomly select one state based on empirical transition probabilities and its length from the empirical distribution of observe lengths of that state.
    \end{enumerate}
\end{enumerate}
}}

\subsubsection*{Paired-MC of higher orders}

Depending on inherent characteristics of the sequences such as dependency between states, we have the capability to implement Markov Chains of higher orders. Take human activity sequences for example: human activity sequences have the nature of trip-activity-trip patterns, which implies that an order of 2 may better estimate dependencies between adjacent states. \cite{song2021visualizing}  Since higher orders often require a higher amount of available data, paired-MC of higher orders stresses the balance between adequately accommodating local patterns and overfitting due to insufficient data. Algorithms for paired-MC of order larger than 2 follow the same process.

\vspace{7mm}

\framebox{\parbox{\dimexpr\linewidth-2\fboxsep-2\fboxrule}{\itshape
{\bf{Algorithm for paired-MC of order 2:} }
\begin{enumerate}
    \item At t = 1, initialize with a state and its length.
    \item Denote the current state as $A_c$ and the end time of $A_c$ as $t_c$. While $t_c < 1440$, repeat:
    \begin{enumerate}
        \item In a certain time interval $\delta$ of $t_c$ (e.g. $\delta = \pm$ 1 hour), find all states with $A_c$ as the preceding state and their lengths. Each sequence may have more than 1 candidate that fulfills the requirement. 
        \item For each potential state, obtain the previous state before $A_c$, denoted as $A_{c-1}$.
        \item Keep potential states whose $A_{c-1}$ matches with $A_{c-1}$ of the sequence being synthesized.
        \item Randomly select one state based on empirical transition probabilities and sample its length from the empirical distribution of lengths.
    \end{enumerate}
\end{enumerate} 
}}

\subsection{Pre-clustering}

An essential component of sequence synthesis is to use accurate transition probabilities when determining the next activities. Sequential categorical data often have inherent subgroup structures. Take human activity sequences as an example: it’s less reasonable to borrow information from non-workdays than workdays if a workday is being synthesized. In addition, we need to have baselines of what realistic sequences look like in order to simulate realistic sequences. This implies that some clustering of sequences exists, so that sequences similar to typical sequences in a cluster are known to be realistic. If the data has little pattern to build meaningful clusters from, it is inherently difficult to judge if it's realistic from this dataset. In addition, synthesis tasks are often related to the subgroup structure of interest. Dividing sequences into clusters offers control and flexibility over the synthesis output. Since clusters can be defined by any common characteristic of the sequences, we can easily "tune the dial" and reweight the clusters to change the composition of output sequences. Such common characteristics may include the source of the sequence and the cluster assignment based on different clustering metrics. 

For the reasons above, we propose pre-clustering data before synthesis and synthesize based on the clustering assignment. Depending on the prior motivating knowledge of the subgroup structure, clustering can be either data-driven or user-driven. Data-driven approaches assume no prior motivation of clustering assignments, and use strategies such as Dunn Index to discover some basic subgroup structure among the sequences. Alternatively, users can pre-specify the cluster assignment if there is a specific driving force behind the clustering assignment (e.g. whether the sequence includes a lunch break) For either approach, borrowing more from “relevant” data within the cluster makes estimations of local transition probabilities more accurate. Below is an implementation example.

\vspace{7mm}

\framebox{\parbox{\dimexpr\linewidth-2\fboxsep-2\fboxrule}{\itshape
{\bf{Algorithm for clustering before synthesis:}}
\vspace{-2mm}
\begin{enumerate}
    \item Specify the cluster assignment
    \begin{itemize}
        \item With prior knowledge, user can input their clustering assignment.
        \item Without prior knowledge, use data-driven clustering approaches. Below is an example for using agglomerative hierarchical clustering:
        \begin{itemize}
            \item Create a pairwise distance matrix for available data used for synthesis
            \item Create dendrogram using hierarchical clustering
            \item To obtain the optimal number of clusters, cut the tree with potential cluster numbers and use the one with the optimal associated Dunn Index.
            \item Identify distinct clusters and group smaller clusters into one large group.
        \end{itemize}
    \end{itemize}
    \item Randomly draw a cluster number with the probability of each cluster as the cluster size divided by the total number of available sequences.
    \item Treat sequences in the chosen cluster as the available data and synthesize the sequence as described above.
\end{enumerate}
}}

\section{Synthesis examples} 

In this section, we show how our algorithm works in practice starting with the motivating example of human activity sequences described in Section 1. To demontrate that our tool is equipped to handle different cases of sequential categorical data, we then move on to a publicly available NHANES accelerometer dataset. For each dataset, we synthesize the same number of sequences as the original dataset using our proposed algorithm and the comparator with and without pre-clustering. We then present summary statistics of synthesized sequences and discuss their performances.

\subsection{Comparisons}

To our knowledge, there has not been an algorithm that offers synthetic data generation for long sequential categorical data due to computational challenges. Time-varying Markov Chains which relies on non-parametric estimation of the transition probability matrix at each time $t$ was used for baseline comparison purposes. The algorithm is as follows: Initialize the sequence by randomly drawing a state at $t = 1$ based on observed probabilities. While the sequence has not been populated (i.e. reach length 1440 for human activity sequences), obtain sequences whose state at time $t-1$ is the same as that of the current sequence, and sample one state as the next state based on empirical transition probabilities. \\

\subsection{Evaluation metrics}

There is no agreed-upon metric for evaluating the quality of sequential categorical data. Transition matrices have been widely regarded as a basic property of categorical sequences, but do not sufficiently account for the time-varying complexity of state transitions. Different variations of statistical similarity and per-instance similarity metrics have been previously proposed on location trace sequences.
\cite{benarous2022synthesis, bindschaedler2016synthesizing, huang2019variational, kulkarni2018generative, rao2020lstm} However, most such metrics do not apply to human activity sequences, either because our data do not directly possess spatial properties such that traces can be visualized on a map or because human activity sequences are much longer. Therefore, we use the following summary statistics to evaluate the match between synthesized and observed human activity sequences:

\begin{enumerate}
    \item Distribution of state frequency, measured by the mean and standard deviation of the number of states in each sequence. 
    \item Conditional distribution of each individual activity-specific duration (i.e. distribution of individual state durations). For five dominant states based on context-specific importance and overall prevalence in the data, we implement visualizations and a Kolmogorov-Smirnov test to demonstrate the similarity between the distributions of state durations from simulated and original sequences. \cite{massey1951kolmogorov} 
    \item Distribution of combined state durations. To this end, we create visualizations and implement the Kolmogorov-Smirnov test to test the similarity between the distributions of combined state lengths from simulated and original sequences.
    \item Distribution of entropies, measured by both the Kolmogorov-Smirnov test and visual inspection of the histogram.
\end{enumerate}

These metrics can be easily modified and adapted to other types of sequential categorical data; for example, for different contexts other summary statistics of individual state distributions and alternative measures of the distance between distributions (e.g., Kullback-Leibler divergence) can be used.

\subsection{Data I: Human activity sequences}

Our first dataset is the motivating data of human activity sequences introduced in Section 1. We synthesize 1929 sequences using approaches described in Section 2: Available sequences are first grouped according to the process laid out in Section 2.4; three clusters of size 911, 808 and 210 were constructed. Computation time is desirable for all proposed approaches: it takes approximately 1 second to synthesize each sequence with parallel computing, given 1929 sequences of original data and 1440 states per sequence. A time window $\delta = 60$ min is used. See the appendix for a complete sensitivity analysis of the choice of different time windows on our algorithm's performance.

Table 2 summarizes the distributions of five largest states' lengths of simulated and original sequences, and compares the distribution between original sequences and each set of simulated sequences using Kolmogorov-Smirnov test. Results show that paired-MC of order 1 and 2 both offer significant improvements over TVMC in terms of the similarity of state distributions between synthesized and original sequences. Using the pre-clustering approach for example, the p-value from a Kolmogorov-Smirnov test of the distribution of home episodes using TVMC is $4 * 10^{-13}$ compared to $0.069$ using paired-MC order 1 and $0.89$ using paired-MC order 2. In the appendix, we further demonstrate scenarios where paired-MC order 2 is superior to paired-MC order 1 in terms of combined state lengths for states $\textit{Leisure and recreation}$ and $\textit{Personal Business}$ (see Table S2).



For summarizing the distributions of several metrics, we obtain the empirical percentile at each position of the sequence, calculate the difference between the percentiles between the original sequences and synthesized sequences at each position, and plot them with GAM smoothing \cite{wood2011fast}. The empirical CDF plots show similar results, where paired-MC curves are closest to the true horizontal line compared to TVMC curves (see Figure 6). Furthermore, pre-clustering is shown to improve the similarity between the entropies of original and synthesized sequences.

\begin{table}[h!]
\centering
\refstepcounter{table}
\tiny
\begin{tblr}{
  cells = {c},
  cell{3}{1} = {r=2}{},
  cell{5}{1} = {r=2}{},
  cell{7}{1} = {r=2}{},
  cell{9}{1} = {r=2}{},
  cell{11}{1} = {r=2}{},
  cell{1}{4} = {c=2}{},
  cell{1}{6} = {c=2}{},
  cell{1}{8} = {c=2}{},
  vline{2-3} = {-}{},
  hline{2-3,5,7,9,11} = {-}{},
}
                          &               & Original     & TVMC          &           & Paired-MC    &         & {Paired-MC order 2} &         \\
                          &               & {Mean (SD)} & {Mean (SD)}  & p-value   & {Mean (SD)} & p-value & {Mean (SD)}          & p-value \\
Home                      & Clustered     & {402 (343)} & {403 (299)}  & $4*10^{-13}$ & {413 (349)} & 0.069   & {398 (334)}          & 0.89    \\
                          & Not clustered &              & {404 (298)}  & $7*10^{-16}$    & {407 (341)} & 0.29    & {403 (347)}          & 0.93    \\
Work                      & Clustered     & {275 (225)} & {283 (222)}  & $3*10^{-4}$    & {277 (226)} & 0.92    & {287 (229)}          & 0.17    \\
                          & Not clustered &              & {269 (215)}  & $3*10^{-7}$    & {280 (224)} & 0.77    & {285 (225)}          & 0.25    \\
Car                       & Clustered     & {16 (15)}   & {16 (16)}    & 0         & {16 (15)}   & 0.68    & {17 (15)}            & 0.80    \\
                          & Not clustered &              & {16 (16)}    & 0         & {16 (15)}   & 0.80    & {16 (15)}            & 0.99    \\
{Leisure and\\recreation} & Clustered     & {112 (173)} & {109 (145)} & $1*10^{-9}$    & {104 (163)} & 0.76    & {117 (172)}          & 0.60    \\
                          & Not clustered &              & {110 (133)}  & $2*10^{-13}$   & {114 (170)} & 0.85    & {114 (168)}          & 0.87    \\
{Personal\\Business}      & Clustered     & {97 (179)}  & {103 (159)}  & $6*10^{-10}$   & {100 (189)} & 0.55    & {107 (193)}          & 0.98    \\
                          & Not clustered &              & {107 (122)}  & 0         & {80 (142)}  & 0.76    & {103 (187)}          & 0.99    
\end{tblr}
\caption{Mean (SD) and p-values of Kolomogorov-Smirnov test between distributions of five largest states' lengths from simulated sequences and original sequences}
\end{table}

\begin{figure}[h!]
\centering
   \includegraphics[width=150mm]{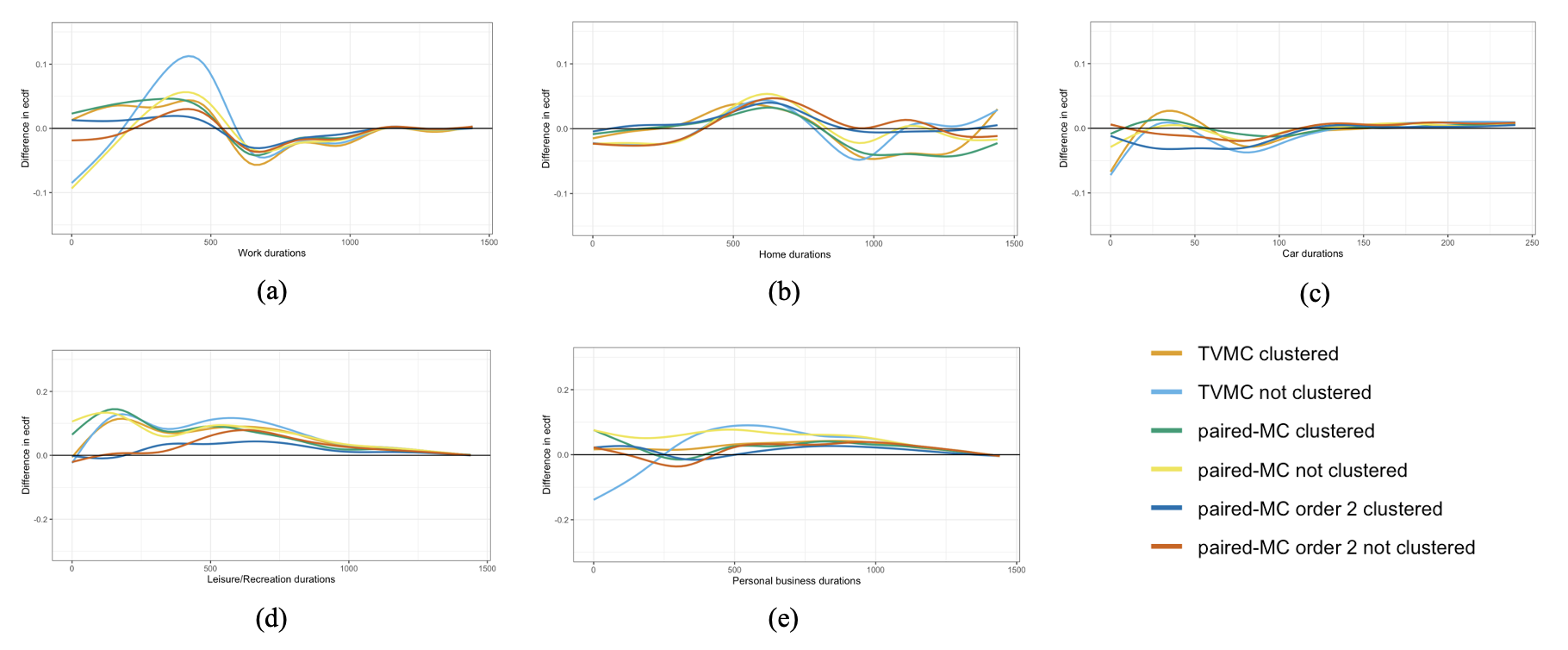}
   \caption{Empirical cdf plot of combined state lengths comparing different synthesis methods for human activity sequence data. From top-left: a) work, b) home, c) car (0-4 hours), d) leisure/recreation (excluding 0 length), e) personal business (excluding 0 length)}
\end{figure}

\begin{figure}[h!]
    \centering
    \includegraphics[width=115mm]{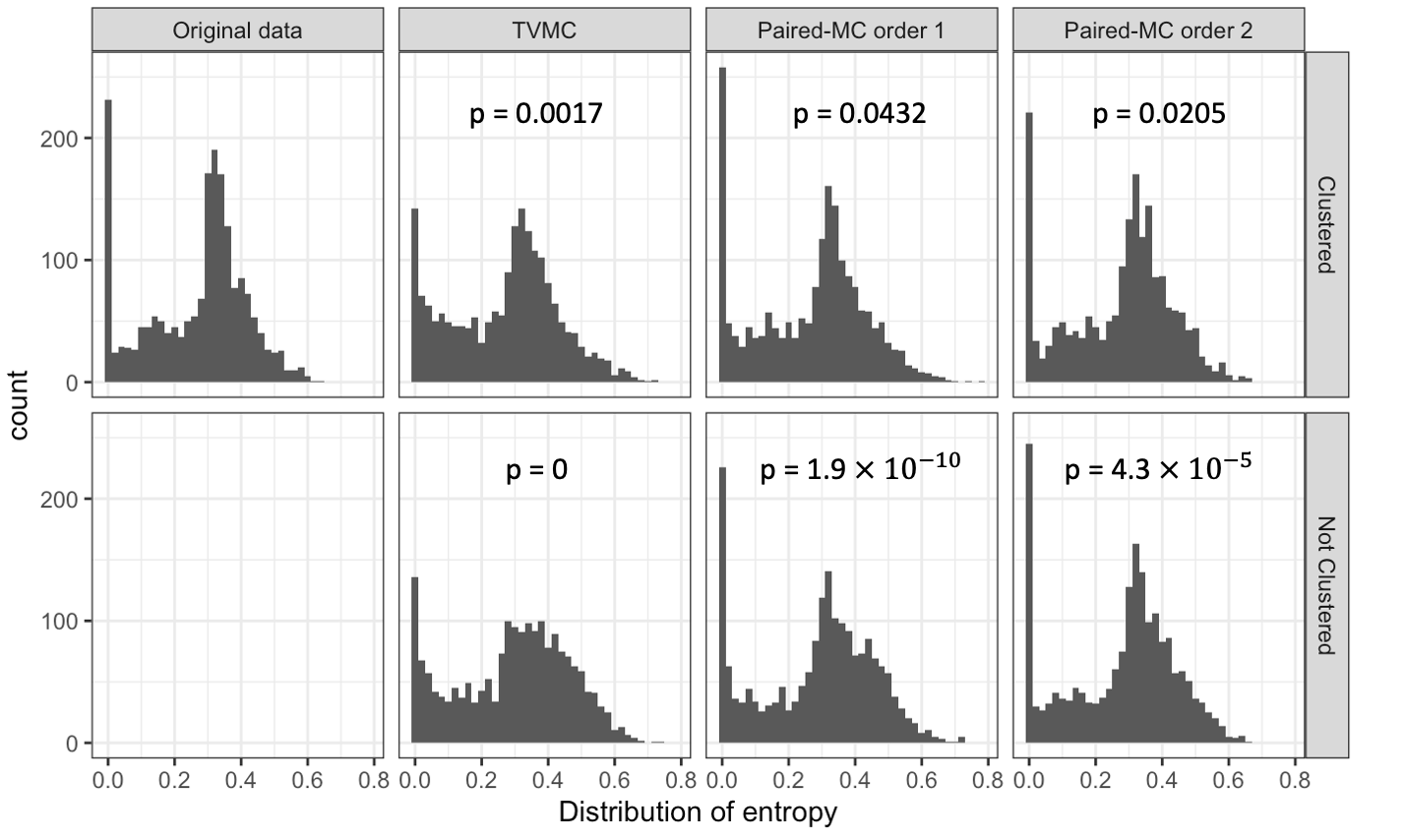}
    \caption{Distribution of entropies and K-S test statistic comparing original sequences with synthesized human activity sequences}
\end{figure}

Table \ref{HAS_number_of_individualstate} summarizes the mean and standard deviation of the number of distinct states in the sequence for the most common activities, which reflects whether the number of transitions is correct and whether transition probabilities are accurate. From the table, we observe that TVMC tends to under-estimate the variability of the number of states for each major activity. Paired-MC order 2 performs relatively better but still under-estimates the variability especially for home episodes. We hypothesize that this may be due to significant outliers in the original data not designed to be picked up by our method. Consider that the mean and standard deviation of the number of "Home" episodes is 2.44 (2.86). However, after removing the two extreme outliers in the number of "Home" episodes: 40 and 107, the standard deviation gets reduced to 1.33, much closer to the target. This shows that our method is better than competing methods in terms of its robustness to outliers.

\begin{table}[h!]
\small
\centering
\caption{Mean (SD) of the number of daily episodes overall and for each major activity in original and synthesized human activity sequences}
\begin{tblr}{
  column{even} = {c},
  column{odd} = {c},
  cell{2}{1} = {r=2}{},
  cell{4}{1} = {r=2}{},
  cell{6}{1} = {r=2}{},
  cell{8}{1} = {r=2}{},
  vline{2} = {1-13}{},
  hline{2,4,6,8,10,12} = {-}{}
  }
Mean (SD) & & Original & TVMC & Paired-MC & Paired-MC order 2 \\
Overall                   & Clustered     & 10.32 (9.04) & 10.14 (5.55) & 10.21 (6.72) & 10.34 (6.46)       \\
                       & Not clustered &             & 10.32 (5.61) & 10.22 (6.26) & 10.19 (6.46)       \\
Home                   & Clustered     & 2.44 (2.86) & 2.44 (1.04) & 2.40 (1.21) & 2.43 (1.27)       \\
                       & Not clustered &             & 2.43 (1.00) & 2.40 (1.14) & 2.41 (1.23)       \\
Work                   & Clustered     & 0.93 (1.39) & 0.90 (1.11) & 0.89 (1.17) & 0.90 (1.23)       \\
                       & Not clustered &             & 0.92 (0.96) & 0.94 (0.97) & 0.92 (1.18)       \\
Car                    & Clustered     & 3.60 (3.92) & 3.56 (2.29) & 3.56 (2.74) & 3.62 (2.58)       \\
                       & Not clustered &             & 3.61 (2.33) & 3.55 (2.53) & 3.54 (2.66)       \\
Leisure and & Clustered     & 0.38 (1.30) & 0.37 (0.73) & 0.36 (0.76) & 0.40 (1.12)       \\
recreation                   & Not clustered &             & 0.38 (0.68) & 0.38 (0.67) & 0.37 (0.96)       \\
Personal      & Clustered     & 0.36 (1.01) & 0.34 (0.69) & 0.37 (0.73) & 0.36 (0.78)       \\
    Business                    & Not clustered &             & 0.37 (0.65) & 0.35 (0.66) & 0.35 (0.78)       
\end{tblr}
\label{HAS_number_of_individualstate}
\end{table}

\subsection{Data II: Accelerometer data}

\subsubsection{Data description and setup}

To demonstrate that our tools can be reasonably implemented in datasets with different underlying data structures, we further consider the NHANES accelerometer data. The NHANES accelerometer data is publicly available on Center for Disease Control and Prevention (downloaded from https://wwwn.cdc.gov/Nchs/Nhanes/2003-2004/PAXRAW\_C.ZIP) and discretized for the purpose of evaluating our method due to its availability. The accelerometer data contains activity counts that were accumulated in bouts of 10 minutes or more (with one or two minute allowed below threshold) for each minute. \cite{boyer2016accelerometer} 2000 daily sequences were first randomly sampled and smoothed with a rolling window of 5 minutes due to the highly variable nature of such data, creating 2000 sequences of length 1436. Thresholds of 760 counts/min and 2020 counts/min are chosen according to \cite{boyer2016accelerometer} as the cut-off for moderate and vigorous activity respectively. Data are thus discretized into four categories: 0 count/min (stationary), 1-760 counts/min, 761-2020 counts/min and $>$ 2020 counts/min. Available sequences were similarly grouped according to the process laid out in Section 2.4; four clusters of size 1060, 491, 230 and 219 were constructed. The data were synthesized according to the paired-MC algorithm outlined in Section 2 with $\delta = 60$ minutes. To our knowledge, such data are not known to possess patterns that allow for an order of higher than 1 (such as the trip-activity-trip pattern for human activity sequences) and therefore paired-MC of order 1 was performed by default.

Results show that our method with pre-clustering offers significant improvement on simulating the distribution of state 1 and 2 lengths (see $p=0.054$ and $p=0.059$ in Table 5). Less improvement is observed for the distributions of States 3 and 4 as well as entropies. We hypothesize that this may be due to the plethora of distinct states even with smoothing, although our algorithm still shows an improvement in terms of the K-S test statistic. Considering that accelerometer data exhibit different time dependence patterns from those in human activity sequences, it may be helpful to conduct a sensitivity analysis of different time window choices. More broadly, sensitivity analyses could be considered for other types of sequential categorical data with many distinct episodes, especially if we have less prior knowledge of the data.

\begin{table}[h!]
\centering
\small
\begin{tblr}{
  row{even} = {c},
  row{odd} = {c},
  cell{2}{1} = {r=2}{},
  cell{4}{1} = {r=2}{},
  cell{6}{1} = {r=2}{},
  cell{8}{1} = {r=2}{},
  vline{2} = {1-11}{},
  hline{2,4,6,8,10} = {-}{},
}
Mean (SD) &               & Original      & TVMC          & paired-MC     \\
Overall   & Clustered     & 59.29 (33.42) & 60.19 (25.29)  & 59.02 (29.42)  \\
          & Not clustered &               & 59.38 (10.45)  & 59.74 (25.93)  \\
State 1   & Clustered     & 14.50 (10.85) & 14.68 (5.31)  & 14.44 (6.82)  \\
          & Not clustered &               & 14.54 (2.87)  & 14.58 (6.57)  \\
State 2   & Clustered     & 26.41 (14.66) & 26.81 (11.00) & 26.35 (13.05) \\
          & Not clustered &               & 26.49 (4.56)  & 26.60 (11.67) \\
State 3   & Clustered     & 15.55 (14.23) & 15.82 (8.75)  & 15.46 (9.40)  \\
          & Not clustered &               & 15.53 (4.80)  & 15.70 (7.83)  \\
State 4   & Clustered     & 2.83 (4.51)   & 2.88 (2.61)   & 2.77 (2.61)   \\
          & Not clustered &               & 2.81 (1.93)   & 2.87 (2.28)   
\end{tblr}
\caption{Mean and standard deviation of the number of daily episodes overall and for each state in original and synthesized accelerometer sequences}
\end{table}

\begin{table}[h!]
\small
\begin{tabular}{c|ccc}
KS-test p-value/test statistic D        &               & TVMC      & paired-MC \\ \hline
\multirow{2}{*}{State 1} & Clustered     & $4.3 \times 10^{-10}$ & 0.054     \\
                         & Not clustered & 0     & $2.5 \times 10^{-11}$ \\ \hline
\multirow{2}{*}{State 2} & Clustered     & $1.4 \times 10^{-13}$ & 0.059     \\
                         & Not clustered & 0         & $1.8 \times 10^{-10}$ \\ \hline
\multirow{2}{*}{State 3 \footnote{\label{note1} P-value outputs of Kolomogorov-Smirnov test for states 3 and 4 durations are 0 for all four sequences, potentially due to the specific data structure. However, paired-MC still performs comparatively better judging by the K-S test statistic D.}} & Clustered     & 0.195     & 0.185     \\
                         & Not clustered & 0.353     & 0.253     \\ \hline
\multirow{2}{*}{$\text{State 4 }^{a}$} & Clustered     & 0.297     & 0.272     \\
                         & Not clustered & 0.383     & 0.318    
\end{tabular}
\caption{p-values of Kolomogorov-Smirnov test between distributions of state 1 and 2 lengths and test statistics of Kolomogorov-Smirnov test between distributions of state 3 and 4 lengths of simulated sequences and original sequences for NHANES accelerometer data}
\end{table}

\begin{figure}[h!]
    \centering
    \includegraphics[width=90mm]{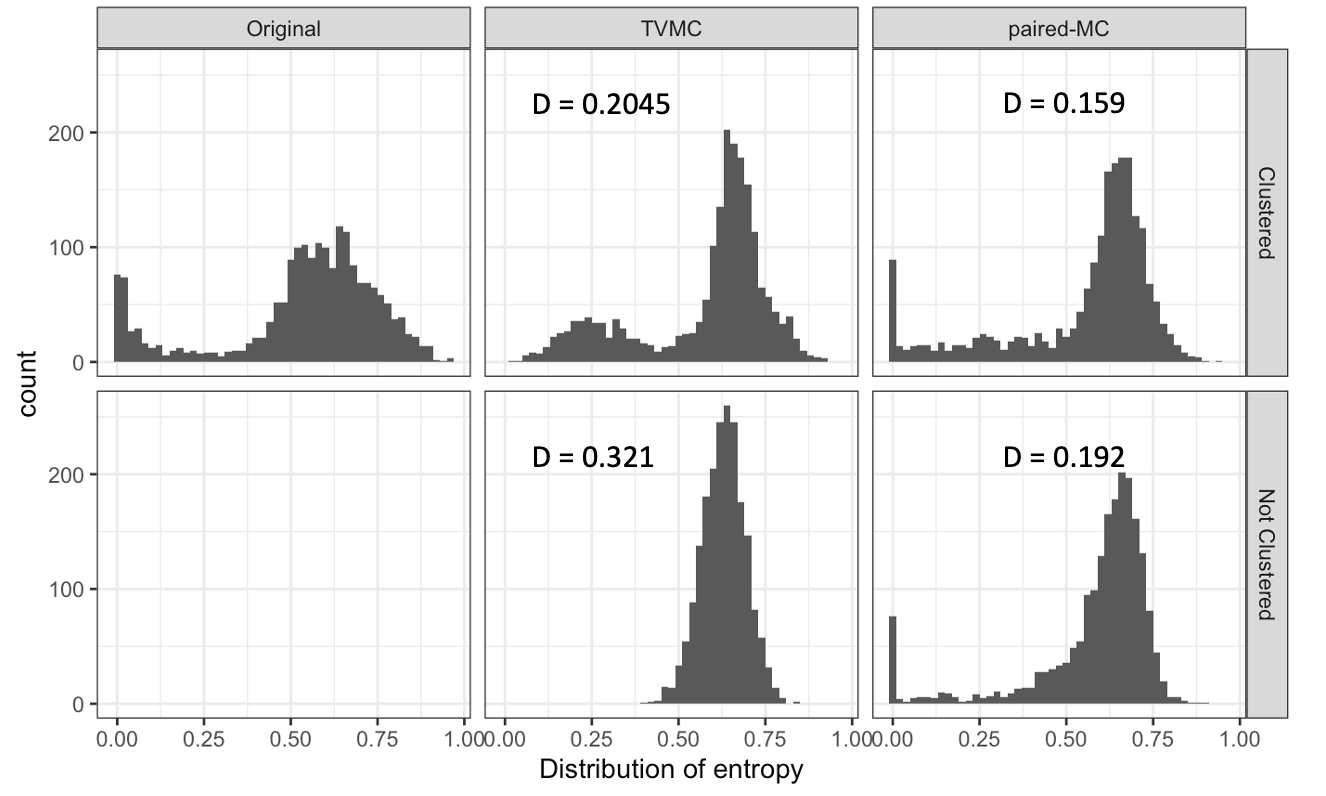}
    \caption{Distributions of entropies and test statistics of Kolomogorov-Smirnov test between distributions of entropies of simulated sequences and original sequences for accelerometer data}
    \footnote{The p-value outputs of Kolomogorov-Smirnov test are 0 for all four sequences, potentially due to the specific data structure. However, paired-MC still performs comparatively better judging by the plot and the K-S test statistic D.}
    \label{entropies_nhanes}
\end{figure}

\newpage

\section{Discussion}

Paired-MC is a novel method used to efficiently synthesize long categorical sequences that are parametrizable and realistic. In contrast to discrete-time Markov Chains, it relaxes the key Markovian property assumption and offers the possibility to use Markov Chains of higher orders on lengthy sequences without sacrificing much computational performance. We further propose pre-clustering as a pre-processing tool that facilitates the estimation of local transition probabilities and offers flexible control over the output sequences. Our examples on human activity sequences and NHANES accelerometer data demonstrate that paired-MC and pre-clustering works on different types of sequential categorical data.  

We note a few limitations of our proposed sequence synthesis algorithm. First, the window size $\delta$ is different for different applications and sequence lengths. A good intuition from domain knowledge or a sensitivity analysis, as shown in Section III of Supplemental Materials, is suggested to accurately estimate transition probabilities. Second, synthesized sequences may underrepresent the variability of source data because the algorithm undersamples "outlying" sequences and is constrained to states and patterns in the observed source data. Third, potential long-range patterns may still be missed in certain applications if the cyclic pattern length exceeds the paired-MC order.

Finally, we discuss some other applications of paired-MC. First, both application examples use sequences of equal lengths in this article. For sequences with varying lengths or survival data such as trajectories of end-stage cancer patients between different stages of care, we anticipate that paired-MC can reasonably accommodate sequences of varying lengths by adding a null/failure state, which is an absorbing state by our algorithm, to the end of shorter sequences. 
Second, entire sequences are generated de novo in both examples. However, partial sequences may be available in some cases, and the goal becomes imputing or predicting the unobserved sequence elements. We plan to investigate the use of paired-MC in these scenarios. Extensions will be studied in future work. Third, synthetic data have long been viewed as a tool to maintain data confidentiality and privacy since Rubin proposed the idea of full synthetic data. \cite{raghunathan2021synthetic, jordon2022synthetic, rubin1993statistical} Active research areas include location-privacy protection mechanisms and disclosure risk measures. \cite{shokri2011quantifying, mcclure2012differential} Privacy is not a concern for our motivating example of human activity data because the activity states types are general and GPS location data is not included. However, many types of categorical sequential data contain sensitive reidentifiable information. It is of research interest to study privacy measures such as reidentification probability for applicable paired-MC output sequences.

\section*{Software}

Software in the form of R code, including datasets and synthesis studies, can be found on Github at https://github.com/zuofuhuang/SimSeq\_Functions.

\section*{Acknowledgment}

Conflict of interest: None declared.

\newpage

\bibliography{bibtex}

\begin{thebibliography}{10}

\bibitem{fan2015smartrac}
Y.~Fan, J.~Wolfson, G.~Adomavicius, K.~Vardhan~Das, Y.~Khandelwal, and J.~Kang,
  ``Smartrac: A smartphone solution for context-aware travel and activity
  capturing,'' 2015.

\bibitem{song2021visualizing}
Y.~Song, S.~Ren, J.~Wolfson, Y.~Zhang, R.~Brown, and Y.~Fan, ``Visualizing,
  clustering, and characterizing activity-trip sequences via weighted sequence
  alignment and functional data analysis,'' {\em Transportation Research Part
  C: Emerging Technologies}, vol.~126, p.~103007, 2021.

\bibitem{dahmen2019synsys}
J.~Dahmen and D.~Cook, ``Synsys: A synthetic data generation system for
  healthcare applications,'' {\em Sensors}, vol.~19, no.~5, p.~1181, 2019.

\bibitem{benarous2022synthesis}
M.~Benarous, E.~Toch, and I.~Ben-Gal, ``Synthesis of longitudinal human
  location sequences: Balancing utility and privacy,'' {\em ACM Transactions on
  Knowledge Discovery from Data (TKDD)}, 2022.

\bibitem{bindschaedler2016synthesizing}
V.~Bindschaedler and R.~Shokri, ``Synthesizing plausible privacy-preserving
  location traces,'' in {\em 2016 IEEE Symposium on Security and Privacy (SP)},
  pp.~546--563, IEEE, 2016.

\bibitem{gabadinho2016analyzing}
A.~Gabadinho and G.~Ritschard, ``Analyzing state sequences with probabilistic
  suffix trees: The pst r package,'' {\em Journal of statistical software},
  vol.~72, pp.~1--39, 2016.

\bibitem{alahi2016social}
A.~Alahi, K.~Goel, V.~Ramanathan, A.~Robicquet, L.~Fei-Fei, and S.~Savarese,
  ``Social lstm: Human trajectory prediction in crowded spaces,'' in {\em
  Proceedings of the IEEE conference on computer vision and pattern
  recognition}, pp.~961--971, 2016.

\bibitem{altche2017lstm}
F.~Altch{\'e} and A.~de~La~Fortelle, ``An lstm network for highway trajectory
  prediction,'' in {\em 2017 IEEE 20th international conference on intelligent
  transportation systems (ITSC)}, pp.~353--359, IEEE, 2017.

\bibitem{parzen1962estimation}
E.~Parzen, ``On estimation of a probability density function and mode,'' {\em
  The annals of mathematical statistics}, vol.~33, no.~3, pp.~1065--1076, 1962.

\bibitem{kooperberg1992logspline}
C.~Kooperberg and C.~J. Stone, ``Logspline density estimation for censored
  data,'' {\em Journal of Computational and Graphical Statistics}, vol.~1,
  no.~4, pp.~301--328, 1992.

\bibitem{stone1997polynomial}
C.~J. Stone, M.~H. Hansen, C.~Kooperberg, and Y.~K. Truong, ``Polynomial
  splines and their tensor products in extended linear modeling: 1994 wald
  memorial lecture,'' {\em The Annals of statistics}, vol.~25, no.~4,
  pp.~1371--1470, 1997.

\bibitem{silverman1986density}
B.~W. Silverman, {\em Density estimation for statistics and data analysis},
  vol.~26.
\newblock CRC press, 1986.

\bibitem{huang2019variational}
D.~Huang, X.~Song, Z.~Fan, R.~Jiang, R.~Shibasaki, Y.~Zhang, H.~Wang, and
  Y.~Kato, ``A variational autoencoder based generative model of urban human
  mobility,'' in {\em 2019 IEEE conference on multimedia information processing
  and retrieval (MIPR)}, pp.~425--430, IEEE, 2019.

\bibitem{kulkarni2018generative}
V.~Kulkarni, N.~Tagasovska, T.~Vatter, and B.~Garbinato, ``Generative models
  for simulating mobility trajectories,'' {\em arXiv preprint
  arXiv:1811.12801}, 2018.

\bibitem{rao2020lstm}
J.~Rao, S.~Gao, Y.~Kang, and Q.~Huang, ``Lstm-trajgan: A deep learning approach
  to trajectory privacy protection,'' {\em arXiv preprint arXiv:2006.10521},
  2020.

\bibitem{massey1951kolmogorov}
F.~J. Massey~Jr, ``The kolmogorov-smirnov test for goodness of fit,'' {\em
  Journal of the American statistical Association}, vol.~46, no.~253,
  pp.~68--78, 1951.

\bibitem{wood2011fast}
S.~N. Wood, ``Fast stable restricted maximum likelihood and marginal likelihood
  estimation of semiparametric generalized linear models,'' {\em Journal of the
  Royal Statistical Society Series B: Statistical Methodology}, vol.~73, no.~1,
  pp.~3--36, 2011.

\bibitem{boyer2016accelerometer}
W.~R. Boyer, D.~L. Wolff-Hughes, D.~R. Bassett, J.~R. Churilla, and E.~C.
  Fitzhugh, ``Accelerometer-derived total activity counts, bouted minutes of
  moderate to vigorous activity, and insulin resistance: Nhanes 2003--2006,''
  2016.

\bibitem{raghunathan2021synthetic}
T.~E. Raghunathan, ``Synthetic data,'' {\em Annual review of statistics and its
  application}, vol.~8, pp.~129--140, 2021.

\bibitem{jordon2022synthetic}
J.~Jordon, L.~Szpruch, F.~Houssiau, M.~Bottarelli, G.~Cherubin, C.~Maple, S.~N.
  Cohen, and A.~Weller, ``Synthetic data--what, why and how?,'' {\em arXiv
  preprint arXiv:2205.03257}, 2022.

\bibitem{rubin1993statistical}
D.~B. Rubin, ``Statistical disclosure limitation,'' {\em Journal of official
  Statistics}, vol.~9, no.~2, pp.~461--468, 1993.

\bibitem{shokri2011quantifying}
R.~Shokri, G.~Theodorakopoulos, J.-Y. Le~Boudec, and J.-P. Hubaux,
  ``Quantifying location privacy,'' in {\em 2011 IEEE symposium on security and
  privacy}, pp.~247--262, IEEE, 2011.

\bibitem{mcclure2012differential}
D.~McClure and J.~P. Reiter, ``Differential privacy and statistical disclosure
  risk measures: An investigation with binary synthetic data.,'' {\em Trans.
  Data Priv.}, vol.~5, no.~3, pp.~535--552, 2012.

\end{thebibliography}

\newpage

\begin{center}
\section*{Supplemental materials}
\end{center}

\setcounter{figure}{0}
\renewcommand{\thefigure}{S\arabic{figure}}

\setcounter{table}{-1}
\renewcommand{\thetable}{S\arabic{table}}

\section*{I. Additional results from human activity sequence data}

\begin{table}[h!]
\centering
\refstepcounter{table}
\tiny
\begin{tblr}{
  cells = {c},
  cell{3}{1} = {r=2}{},
  cell{5}{1} = {r=2}{},
  cell{7}{1} = {r=2}{},
  cell{9}{1} = {r=2}{},
  cell{11}{1} = {r=2}{},
  cell{13}{1} = {r=2}{},
  cell{15}{1} = {r=2}{},
  cell{17}{1} = {r=2}{},
  cell{19}{1} = {r=2}{},
  cell{21}{1} = {r=2}{},
  cell{23}{1} = {r=2}{},
  cell{25}{1} = {r=2}{},
  cell{27}{1} = {r=2}{},
  cell{1}{4} = {c=2}{},
  cell{1}{6} = {c=2}{},
  cell{1}{8} = {c=2}{},
  vline{2-3} = {-}{},
  hline{2-3,5,7,9,11,13,15,17,19,21,23,25,27} = {-}{},
}
                          &               & Original     & TVMC          &           & Paired-MC    &         & {Paired-MC order 2} &         \\
                          &               & {Mean (SD)} & {Mean (SD)}  & p-value   & {Mean (SD)} & p-value & {Mean (SD)}          & p-value \\
Home                      & Clustered     & {402 (343)} & {403 (299)}  & $4*10^{-13}$ & {413 (349)} & 0.069   & {398 (334)}          & 0.89    \\
                          & Not clustered &              & {404 (298)}  & $7*10^{-16}$    & {407 (341)} & 0.29    & {403 (347)}          & 0.93    \\
Work                      & Clustered     & {275 (225)} & {283 (222)}  & $3*10^{-4}$    & {277 (226)} & 0.92    & {287 (229)}          & 0.17    \\
                          & Not clustered &              & {269 (215)}  & $3*10^{-7}$    & {280 (224)} & 0.77    & {285 (225)}          & 0.25    \\
Car                       & Clustered     & {16 (15)}   & {16 (16)}    & 0         & {16 (15)}   & 0.68    & {17 (15)}            & 0.80    \\
                          & Not clustered &              & {16 (16)}    & 0         & {16 (15)}   & 0.80    & {16 (15)}            & 0.99    \\
{Leisure and\\recreation} & Clustered     & {112 (173)} & {109 (145)} & $1*10^{-9}$    & {104 (163)} & 0.76    & {117 (172)}          & 0.60    \\
                          & Not clustered &              & {110 (133)}  & $2*10^{-13}$   & {114 (170)} & 0.85    & {114 (168)}          & 0.87    \\
{Personal\\Business}      & Clustered     & {97 (179)}  & {103 (159)}  & $6*10^{-10}$   & {100 (189)} & 0.55    & {107 (193)}          & 0.98    \\
                          & Not clustered &              & {107 (122)}  & 0         & {80 (142)}  & 0.76    & {103 (187)}          & 0.99 \\
{Bike}      & Clustered     & {15 (16)}  & {16 (20)}  & 0.99   & {15 (15)} & 1    & {12 (12)}          & 0.30    \\
                          & Not clustered &              & {12 (14)}  & 0.30         & {13 (10)}  & 0.60    & {17 (19)}          & 0.91 \\
{Bus}      & Clustered     & {9 (5)}  & {9 (5)}  & 0.51   & {8 (2)} & 0.84    & {9 (3)}          & 0.62    \\
                          & Not clustered &              & {7 (3)}  & 0.51         & {9 (4)}  & 0.73    & {9 (5)}          & 0.63 \\
{Eat out}      & Clustered     & {33 (37)}  & {34 (34)}  & 0.001   & {33 (36)} & 1    & {36 (39)}          & 0.95    \\
                          & Not clustered &              & {33 (34)}  & $4 * 10^{-4}$         & {30 (34)}  & 0.95    & {34 (36)}          & 0.81 \\
{Education}      & Clustered     & {30 (71)}  & {18 (34)}  & 0.70   & {21 (59)} & 0.67    & {28 (67)}          & 0.99    \\
                          & Not clustered &              & {40 (80)}  & 0.60         & {16 (32)}  & 0.09   & {30 (65)}          & 0.90 \\
{N/A}      & Clustered     & {112 (111)}  & {106 (103)}  & 0.014   & {124 (111)} & 0.42    & {107 (110)}          & 0.97    \\
                          & Not clustered &              & {118 (111)}  & $3 * 10^{-4}$        & {123 (117)}  & 0.44    & {112 (111)}          & 0.82 \\
{Other}      & Clustered     & {35 (61)}  & {35 (44)}  & 0   & {33 (58)} & 1    & {37 (65)}          & 0.74    \\
                          & Not clustered &              & {36 (38)}  & 0         & {34 (65)}  & 0.94    & {38 (72)}          & 0.97 \\
{Shop}      & Clustered     & {23 (27)}  & {24 (25)}  & 0.38   & {23 (27)} & 0.74    & {24 (31)}          & 1    \\
                          & Not clustered &              & {23 (22)}  & 0.67         & {22 (27)}  & 0.91    & {22 (23)}          & 1 \\
{Walk}      & Clustered     & {9 (10)}  & {9 (9)}  & $7 * 10^{-13}$   & {9 (11)} & 1    & {9 (11)}          & 1    \\
                          & Not clustered &              & {9 (9)}  & $6 * 10^{-16}$         & {9 (10)}  & 1   & {9 (10)}          & 1

\end{tblr}
\caption{Mean (SD) and p-values of Kolomogorov-Smirnov test between distributions of all states' lengths from simulated sequences and original sequences}
\end{table}

\begin{figure}[h!]
\centering
   \includegraphics[width=150mm]{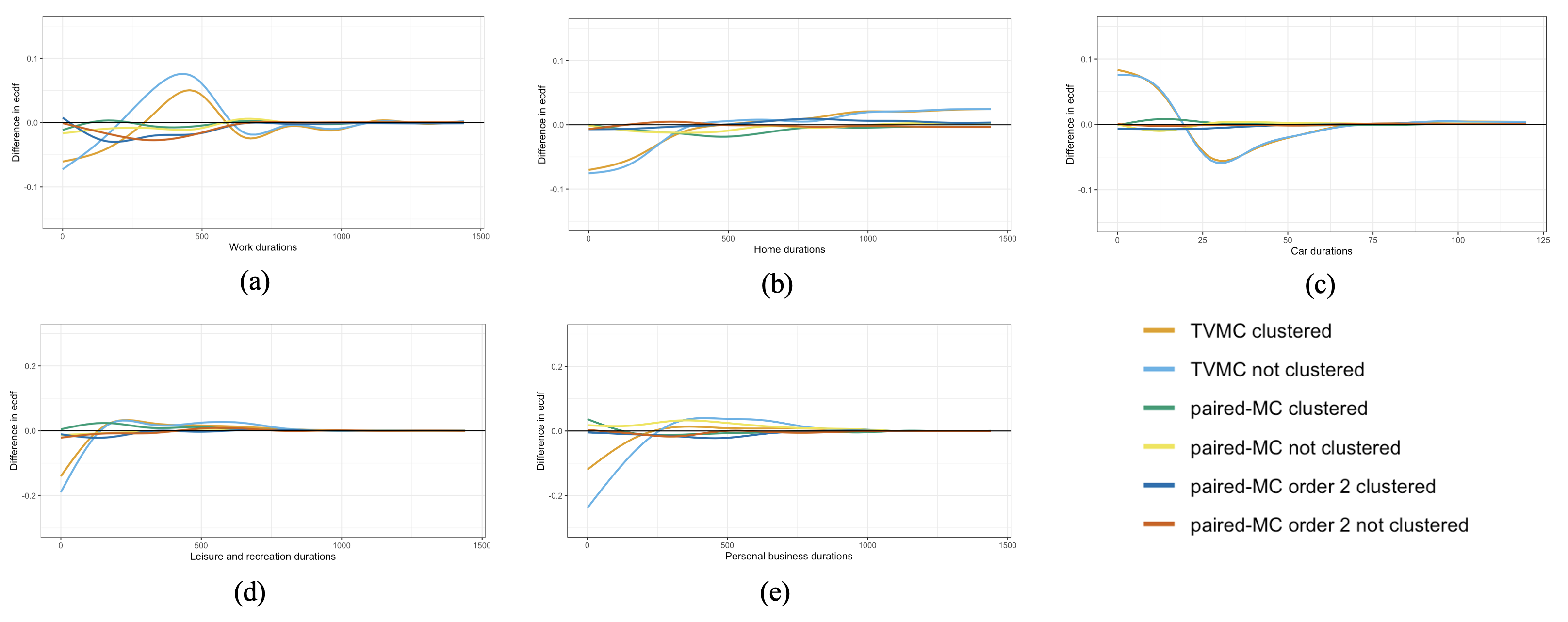}
   \caption{Empirical cdf plot of individual state lengths comparing different synthesis methods for human activity sequence data. From top-left: a) work, b) home, c) car (0-4 hours), d) leisure/recreation (excluding 0 length), e) personal business (excluding 0 length)}
\end{figure}

\begin{table}[h!]
\centering
\caption{p-values from Kolmogorov-Smirnov tests of combined state lengths of synthesized human activity sequences}
\begin{tblr}{
  cells = {c},
  cell{2}{1} = {r=2}{},
  cell{4}{1} = {r=2}{},
  cell{6}{1} = {r=2}{},
  cell{8}{1} = {r=2}{},
  cell{10}{1} = {r=2}{},
  vline{2} = {1-11}{},
  hline{2,4,6,8,10} = {-}{},
}
K-S test p-value       &               & TVMC      & paired-MC & paired-MC order 2 \\
Work                   & Clustered     & 0.002     & 0.019     & 0.12             \\
                       & Not clustered & $3.5*10^{-13}$ & $7.8*10^{-11}$ & 0.11             \\
Home                   & Clustered     & 0.002     & 0.023     & 0.056             \\
                       & Not clustered & 0.001     & 0.008     & 0.014             \\
Car                    & Clustered     & $3.2*10^{-5}$  & 0.83      & 0.027             \\
                       & Not clustered & $2.7*10^{-5}$  & 0.51      & 0.59              \\
Leisure and recreation & Clustered     & $5.0*10^{-5}$  & $8.9*10^{-4}$  & 0.072             \\
                       & Not clustered & $5.0*10^{-11}$ & $6.3*10^{-11}$ & 0.14              \\
Personal Business      & Clustered     & 0.004     & 0.006     & 0.079             \\
                       & Not clustered & $4.5*10^{-10}$ & $2.7*10^{-4}$    & 0.41              \\
\end{tblr}
\label{HAS_KStest_combined_length}
\end{table}

\newpage

\section*{II: Additional results from NHANES accelerometer data}

\begin{table}[h!]
\centering
\small
\caption{Mean and standard deviation of state durations for acelerometer data}
\begin{tabular}{cccccc}
\begin{tabular}[c]{@{}c@{}}Mean\\(SD)\end{tabular} & \begin{tabular}[c]{@{}c@{}}Original\\sequences\end{tabular} & \begin{tabular}[c]{@{}c@{}}TVMC\\Clustered\end{tabular}  & \begin{tabular}[c]{@{}c@{}}TVMC\\Not clustered\end{tabular} & \begin{tabular}[c]{@{}c@{}}Paired-MC \\Clustered\end{tabular} & \begin{tabular}[c]{@{}c@{}}Paired-MC \\Not clustered\end{tabular}  \\ 
\hline
State 1                                            & \begin{tabular}[c]{@{}c@{}}58.15~\\(159.98)\end{tabular}    & \begin{tabular}[c]{@{}c@{}}56.88~\\(110.50)\end{tabular} & \begin{tabular}[c]{@{}c@{}}57.86~\\(82.72)\end{tabular}     & \begin{tabular}[c]{@{}c@{}}58.30\\~(161.21)\end{tabular}      & \begin{tabular}[c]{@{}c@{}}57.77~\\(158.83)\end{tabular}           \\ 
\hline
State 2                                            & \begin{tabular}[c]{@{}c@{}}18.93\\(27.48)\end{tabular}      & \begin{tabular}[c]{@{}c@{}}18.87~\\(19.20)\end{tabular}  & \begin{tabular}[c]{@{}c@{}}18.95\\(18.77)\end{tabular}      & \begin{tabular}[c]{@{}c@{}}18.97\\(27.44)\end{tabular}        & \begin{tabular}[c]{@{}c@{}}18.76~\\(26.84)\end{tabular}            \\ 
\hline
State 3                                            & \begin{tabular}[c]{@{}c@{}}4.68~\\(4.76)\end{tabular}       & \begin{tabular}[c]{@{}c@{}}4.71~\\(4.24)\end{tabular}    & \begin{tabular}[c]{@{}c@{}}4.69~\\(4.15)\end{tabular}       & \begin{tabular}[c]{@{}c@{}}4.66~\\(4.41)\end{tabular}         & \begin{tabular}[c]{@{}c@{}}4.70\\(4.45)\end{tabular}               \\ 
\hline
State 4                                            & \begin{tabular}[c]{@{}c@{}}7.14\\(34.48)\end{tabular}       & \begin{tabular}[c]{@{}c@{}}7.16\\(7.70)\end{tabular}     & \begin{tabular}[c]{@{}c@{}}7.01~\\(7.05)\end{tabular}       & \begin{tabular}[c]{@{}c@{}}8.17~\\(48.66)\end{tabular}        & \begin{tabular}[c]{@{}c@{}}7.19~\\(34.76)\end{tabular}             \\
\hline
\end{tabular}
\end{table}

\begin{table}[h!]
\small
\begin{tabular}{c|ccc}
KS-test p-value        &               & TVMC      & paired-MC \\ \hline
\multirow{2}{*}{State 1} & Clustered     & 0 & 1     \\
                         & Not clustered & 0     & 1 \\ \hline
\multirow{2}{*}{State 2} & Clustered     & 0 & 0.96     \\
                         & Not clustered & 0         & 1 \\ \hline
\multirow{2}{*}{State 3} & Clustered     & 0     & 1     \\
                         & Not clustered & 0     & 1     \\ \hline
\multirow{2}{*}{State 4} & Clustered     & 0     & 0.99     \\
                         & Not clustered & 0     & 1    
\end{tabular}
\caption{p-values of Kolomogorov-Smirnov test between distributions of individual lengths for all states}
\end{table}

\begin{figure}[h!]
    \centering
    \begin{multicols}{2}[]
        \includegraphics[width=80mm]{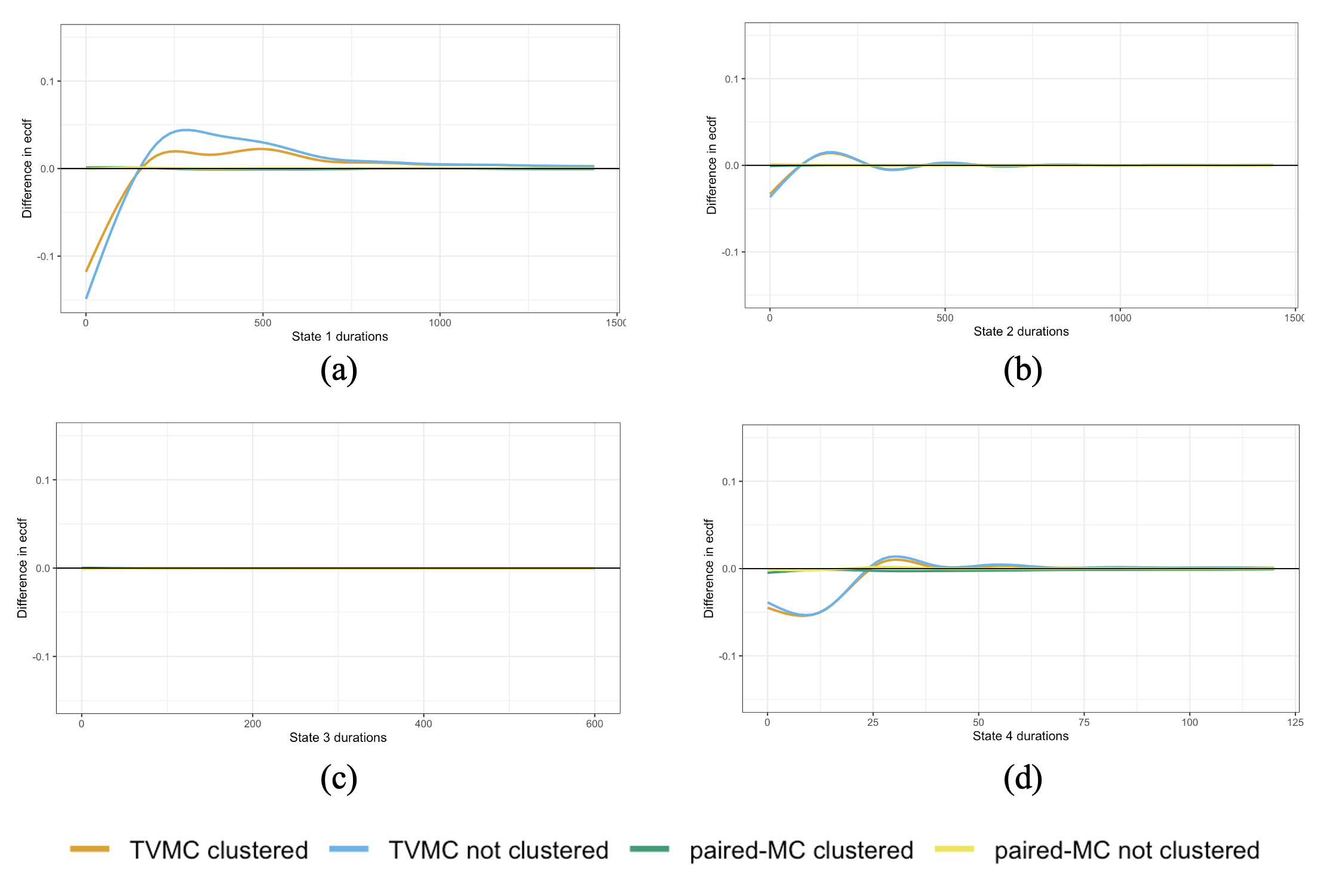}
        \includegraphics[width=80mm]{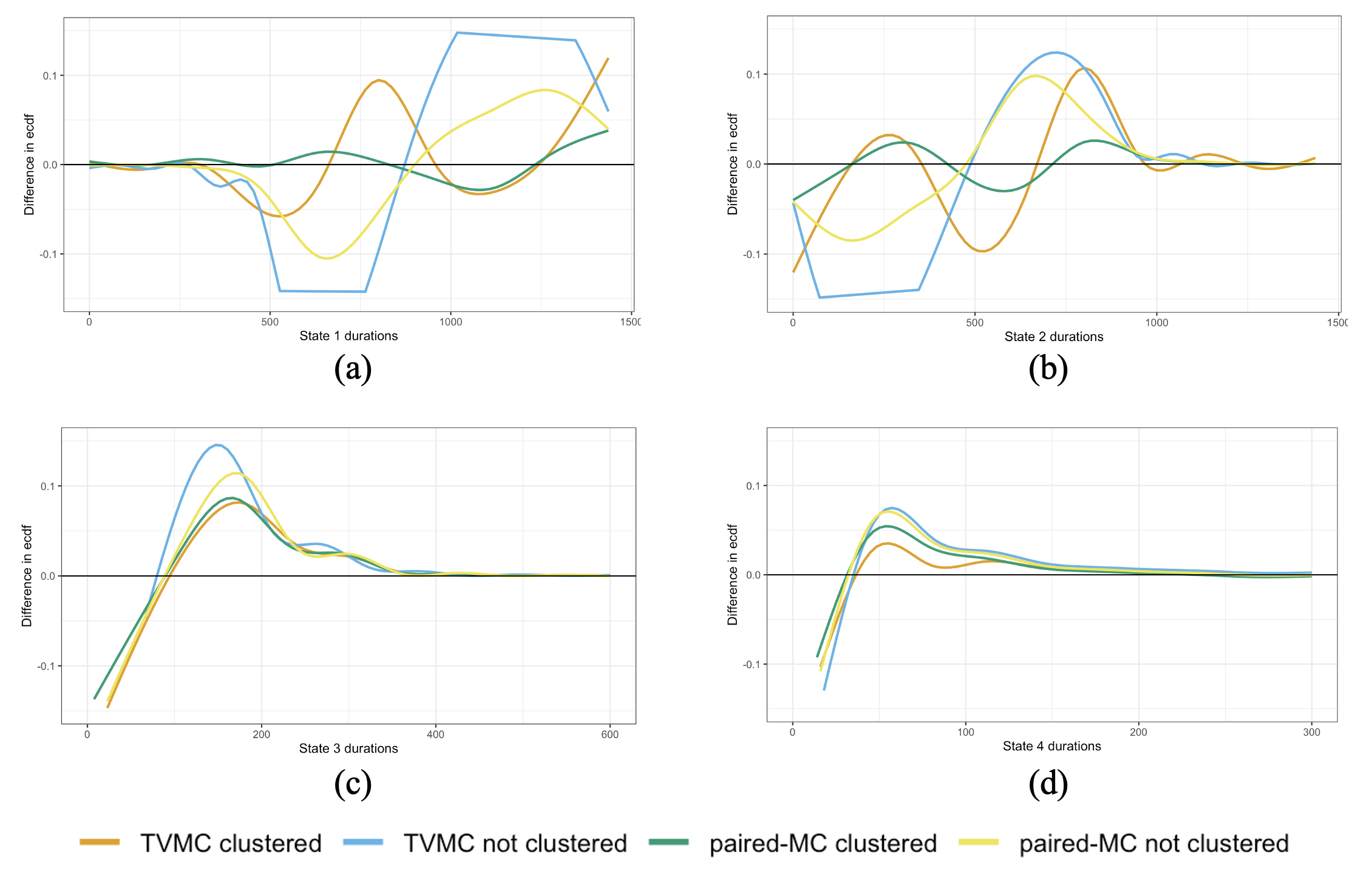}
    \end{multicols}
    \caption{Empirical cdf plots of NHANES accelerometer data. Left panel: individual state length; right panel: combined state length. From top-left: a) State 1, b) State 2, c) State 3 (0-10 hours), d) State 4 (0-5 hours)}
\end{figure}

\newpage

\section*{III: Sensitivity analysis of human activity sequence data}

Sections 3 and 4 demonstrate the superiority of paired-MC with pre-clustering. In this section, we perform a sensitivity analysis that explores the effect of time window $\delta$ on the synthesis results. We further hypothesize that the appropriate window size $\delta$ depends on the nature of the data being synthesized. Next, we perform a sensitivity analysis of the different choices of $\delta$ on the synthesis results. Continue our example of daily trajectories of healthcare workers during COVID-19, where $\delta = 30$ min and $\delta = 120$ min are considered in addition to $\delta = 60$ min. From K-S test results of the distributions of entropies, there is no statistically significant difference in distributions of entropies between original sequences and synthesized sequences using paired-MC with $\delta = 30$. This indicates that the next predicted activity at a certain time point is adequately informed by the patterns within a 30-minute time window for our data size. From distributions of state durations, we obsereve that car episodes are not sensitive to time window due to its prevalence throughout the day. In addition, it is reinforced that less frequent states (such as leisure/recreation and personal business) are sensitive to trip-activity-trip patterns and thus benefit from order 2 estimation. In conclusion, human activity sequences follow a trip-activity-trip pattern and is best approximated by an order-2 paired-MC.

\begin{table}[h!]
\centering
\small
\begin{tblr}{
  row{1} = {c},
  row{2} = {c},
  cell{3}{1} = {c},
  cell{3}{3} = {c},
  cell{3}{4} = {c},
  cell{3}{5} = {c},
  vline{2} = {-}{},
  hline{2-3} = {-}{},
}
{Mean \\(SD)}      & {Original\\sequences} & {paired-MC \\ $\delta = 60$ min}            & {paired-MC \\ $\delta = 30$ min}   & {paired-MC \\ $\delta = 120$ min}  \\
Order 1     & {10.32\\(9.04)}       & {10.21\\(6.72)} & {10.23\\(6.38)}      & {10.28\\(6.61)}      \\
Order 2 &                       & {10.34\\(6.46)} & {10.30\\(6.89)}      & {10.21\\(6.54)}      
\end{tblr}
\caption{Mean and standard deviation of number of states daily}
\end{table}

\begin{figure}[h!]
    \centering
    \includegraphics[width=90mm]{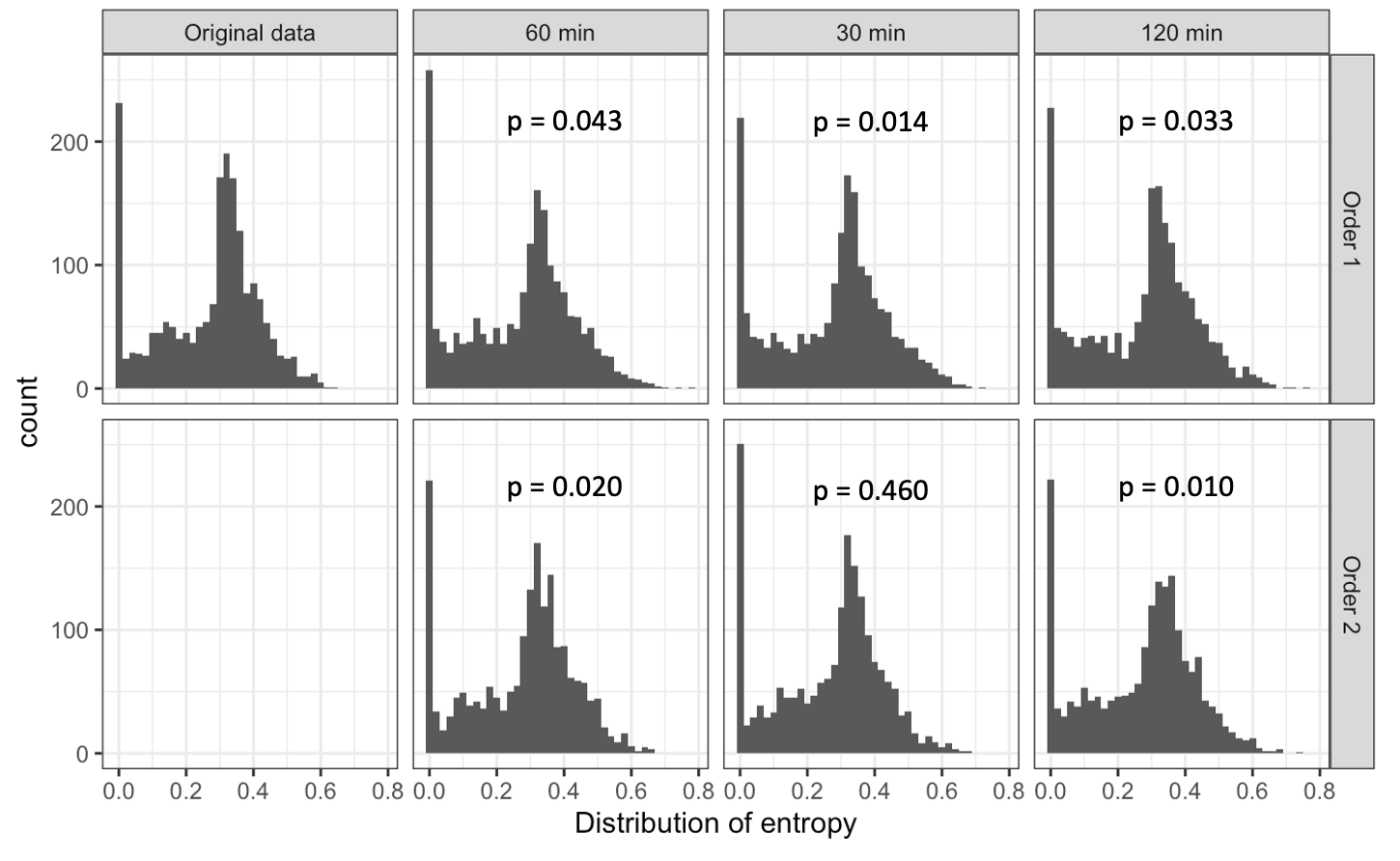}
    \caption{Distributions of entropies and p-values of Kolomogorov-Smirnov test between distributions of entropies of simulated sequences and original sequences}
    \label{entropies_dist_window}
\end{figure}

\begin{table}[h!]
\small
\begin{tabular}{c|cccc}
KS-test p-value                         &         & \begin{tabular}[c]{@{}c@{}}paired-MC\\ $\delta$ = 30 min\end{tabular} & \begin{tabular}[c]{@{}c@{}}paired-MC\\ $\delta$ = 60 min\end{tabular} & \begin{tabular}[c]{@{}c@{}}paired-MC\\ $\delta$ = 120 min\end{tabular} \\ \hline
\multirow{2}{*}{Work}                   & Order 1 & 0.004                                                             & 0.018                                                             & 0.009                                                              \\
                                        & Order 2 & 0.085                                                             & 0.117                                                             & 0.009                                                              \\ \hline
\multirow{2}{*}{Home}                   & Order 1 & 0.005                                                             & 0.023                                                             & 0.009                                                              \\
                                        & Order 2 & 0.067                                                             & 0.056                                                             & 0.085                                                              \\ \hline
\multirow{2}{*}{Car}                    & Order 1 & 0.328                                                             & 0.825                                                             & 0.536                                                              \\
                                        & Order 2 & 0.616                                                             & 0.027                                                             & 0.909                                                              \\ \hline
\multirow{2}{*}{Leisure and recreation} & Order 1 & $5*10^{-5}$                                                            & 0.001                                                             & $5*10^{-5}$                                                             \\
                                        & Order 2 & 0.460                                                             & 0.072                                                             & 0.992                                                              \\ \hline
\multirow{2}{*}{Personal Business}      & Order 1 & 0.023                                                             & 0.006                                                             & 0.036                                                              \\
                                        & Order 2 & 0.169                                                             & 0.079                                                             & 0.724                                                             
\end{tabular}
\caption{p-values of Kolomogorov-Smirnov test between distributions of daily combined length of five major states from simulated sequences and original sequences}
\end{table}

\begin{figure}[h!]
   \centering
   \includegraphics[width=90mm]{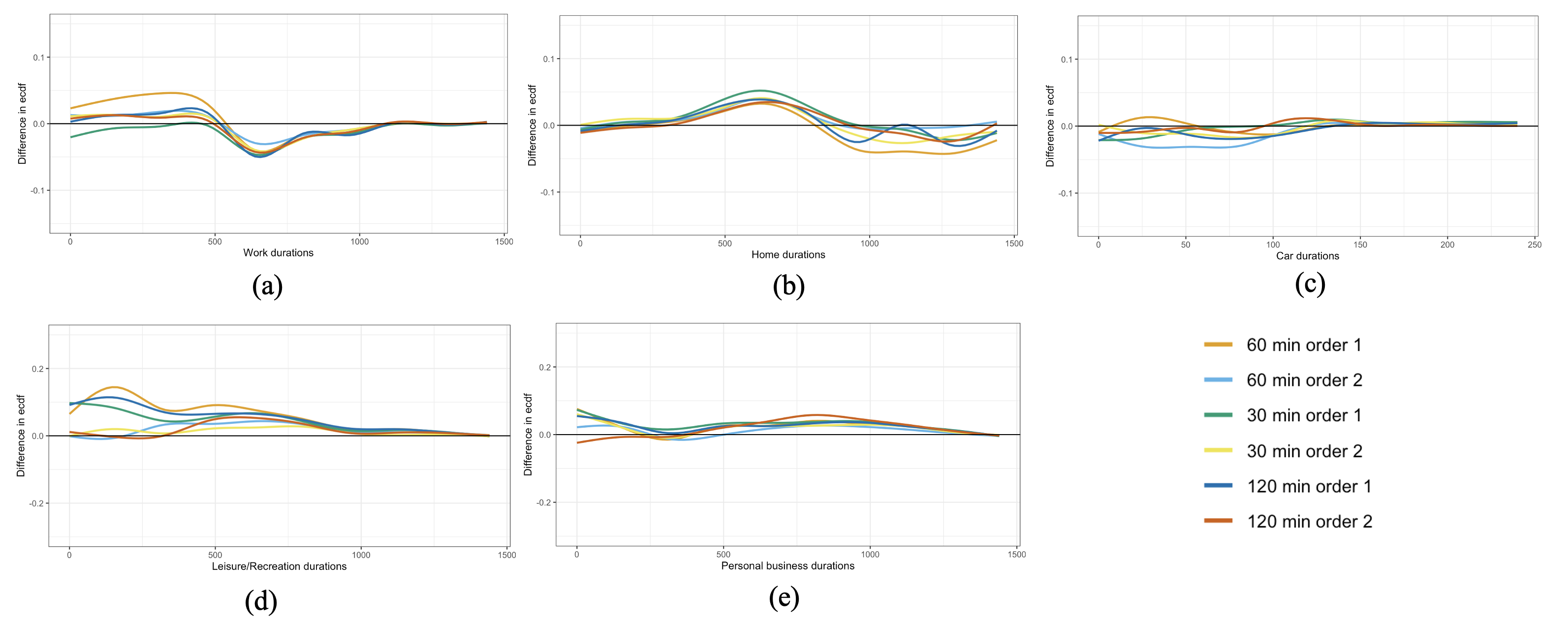}
   \caption{Empirical cdf plots of combined state lengths comparing different $\delta$ values for implementing paired-MC on human activity sequence data a) work, b) home, c) car (0-4 hours), d) leisure/recreation (excluding 0 length), e) personal business (excluding 0 length)}
\end{figure}

\end{document}